\documentstyle[prl,aps]{revtex}
\begin{document}

\title{Phonon-assisted luminescence of magnetoexcitons in
 semiconductor quantum  wells} 

\author
{V.M. Apalkov and M.E. Portnoi
\footnote{Corresponding author: Tel: +44-1392-264154; 
 Fax: +44-1392-264111; E-mail address: m.e.portnoi@ex.ac.uk}}
\address
{School of Physics, University of Exeter, Stocker Road, 
Exeter EX4 4QL, United Kingdom}

\maketitle

\begin{abstract}
We consider a line-shape of magnetoexciton photoluminescence
from quantum wells when the disorder is sufficiently small. 
In this case the phonon-assisted optical transitions become 
important for the line formation. We study both inter-band and 
intra-band excitons. For inter-band excitons the width of a 
single peak emission line is calculated as a function of 
temperature and quantum well width. For intra-band excitons the
double peak of the emission line is predicted when the electron 
filling factor is odd and greater or equal to three. In the latter 
case the lowest magnetoexciton dispersion curve has a minimum
at non-zero momentum. Then the higher-energy  peak results from 
the direct optical emission of zero-momentum excitons. 
The origin of the lower-energy peak is the phonon-assisted 
transitions from the non-zero momentum exciton states. 
With increasing temperature, the higher-energy peak 
becomes more pronounced and the lower-energy peak vanishes.\\

\end{abstract}

 A two-dimensional (2D) exciton in a strong magnetic field is optically 
 active only when the electron and the hole of the exciton belong 
 either to the same Landau level of the different subbands or to 
 the adjacent Landau levels of the same subband. In both cases the 
 optical exciton recombination is allowed only at zero 2D momentum of 
 the exciton. Then the exciton emission consists of a single infinitely 
 narrow line.   
 To acquire the finite width of the line the exciton should transfer 
 its momentum to a third body, which can be either a phonon or an 
 impurity. In this paper we study the phonon-assistant exciton 
 recombination, which can be important in pure enough samples. 
 The material parameters used in this paper correspond to a 
 GaAs quantum well. 

 To find the line-shape of the exciton photoluminescence we restrict 
 ourselves to the  second order phonon-photon processes. 
The electron-phonon Hamiltonian can be rewritten as a sum of 
the electron-phonon and hole-phonon contributions: 
\begin{eqnarray}
H_{e,h-pn} & = & - \sum _{j, \vec{Q}} \frac{M_{e,j} (\vec{Q})}{\sqrt{V}} 
  Z_e(q_z) \left[ 
   \hat{\rho }_{e}^{+}(\vec{q}) \hat{d}^{+}_j (\vec{Q}) + 
   \hat{\rho }_{e}(\vec{q}) \hat{d} _j (\vec{Q}) \right]         \nonumber  \\
      &  &     - \sum _{j, \vec{Q}} \frac{M_{h,j} (\vec{Q})}{\sqrt{V}} 
     Z_h(q_z) \left[ 
   \hat{\rho }_{h}^{+}(\vec{q}) \hat{d}^{+}_j (\vec{Q}) + 
   \hat{\rho }_{h}(\vec{q}) \hat{d} _j (\vec{Q}) \right] \mbox{\hspace{3mm},}
\end{eqnarray}
 where the capital letter ($\vec{Q}$) denotes the three dimensional 
 (3D) vector, its projections are denoted by the corresponding small letters, 
 $\vec{Q}=(\vec{q},q_z)$; $j$ is labeling the phonon modes, 
 $j=1$ for the longitudinal mode  
 and $j=2,3$ for two transverse modes,
 $\hat{d}^{+}_j,\hat{d}_j$ are the creation and annihilation operators
 of a phonon in the $j$th mode, $V$ is a normalization volume; 
 $M_{e,j} (\vec{Q})$ and $M_{h,j} (\vec{Q})$ are the matrix elements  
 of electron-phonon and hole-phonon 
 interactions, which are determined by the deformation potential and 
 piezoelectric coupling. In GaAs they have the form \cite{benedict}:
\begin{equation}
\left\{ \begin{array}{c} M_{e,j} (\vec{Q}) \\ M_{h,j} (\vec{Q}) 
 \end{array} \right\}
   = \sqrt{\frac {\hbar }{2 \rho_0 s Q}} 
      \left[ \left\{ \begin{array}{c} - \\ + \end{array} \right\}  
           \frac{eh_{14}}{\kappa \epsilon _0} 
           \frac{Q_x Q_y \xi _{j,z} +Q_y Q_z \xi _{j,x} +
                           Q_z Q_x \xi _{j,y}}{Q^2}
    - i \left\{ \begin{array}{c} \Xi _e \\ \Xi _h \end{array} \right\}
         (\vec{\xi}_{j} \cdot \vec{Q} ) \right] \mbox{,}
\end{equation}
 where $\rho _0 $ is the mass density of GaAs, 
 $h_{14}$, $\Xi _e$, and $\Xi _h$  are the parameters of piezoelectric
 and deformation potential couplings, $\vec{\xi}_j$ is the polarization
 vector of the $j$th phonon mode. In the above expressions we have 
 used the isotropic Debye approximation 
 with a linear dependence of the phonon frequency on the wave vector:
$\omega _{j} (Q) = s_{j} Q $,
where $s_{j}$ is the speed of sound of the $j$th mode.

 The form factors $Z_{e}(q_z)$ and $Z_{h}(q_z)$ in equation (1) are 
 determined by the electron and hole 
 spreading in $z$ direction and are given by the expression:
\begin{equation}
\left\{ \begin{array}{c} Z_{e}(q_z)  \\ Z_{h}(q_z) \end{array} \right\}
 = \int dz e^{i q_z z} 
\left\{ \begin{array}{c} \left| \chi_e(z) \right|^2   \\ \left| 
 \chi_h(z) \right|^2  \end{array} \right\}  \mbox{\hspace{3mm},}
\end{equation}
where $\chi _e(z)$ and $\chi _h(z)$ are the wave functions  associated 
with the electron and hole subbands, respectively. 
 Bellow we use the infinite well potential with the same width $a$  
 for both the electron and the hole.

 In the second order photon-phonon transitions there are two
 processes which contribute  into the exciton emission distribution.
 In the first process the exciton with the momentum $\vec{k}$ makes the 
 transition into a zero-momentum state with emission of a 
 phonon with momentum  $\vec{Q}$. Then the zero-momentum 
 exciton emits the photon with the energy: $\hbar \omega = 
 E(k) - \hbar \omega (Q)+const$, where $E(k)$ is the binding 
 energy of an exciton with the  momentum $\vec{k}$; the 
 $const$ includes the interband gap energy, for convenience 
 we put this constant to zero. In the second process the
 exciton with the momentum $\vec{k}$ absorbs the phonon with the momentum 
 $\vec{Q}$ and then emits the photon with the energy 
$\hbar \omega =  E(k) + \hbar \omega (Q)$. Taking into 
 account both these processes we find the probability of 
 emitting a photon with frequency $\omega $
\begin{eqnarray}
I(\omega ) &  = & \frac{2\pi }{\hbar} \sum_{\vec{q}} f(q)\sum _{\vec{Q}} 
 \left[   \frac{ \left| < \vec{q}, N(-\vec{Q})+1| H_{e,h-pn}| N(-\vec{Q}),0 > 
\right|^2 }
 { \hbar ^2(\Delta \omega )^2 } 
 \delta (\hbar \Delta \omega - \Delta E(q) - \hbar \omega(Q))
        \right.
                                               \nonumber \\
 &  &  \mbox{\hspace{-17mm}}
  +  \left.  
 \frac{ \left| < \vec{q}, N(\vec{Q})| H_{e,h-pn}| N(\vec{Q})+1,0 > \right|^2 }
  { \hbar ^2(\Delta \omega) ^2 } 
     \delta (\hbar \Delta \omega  - \Delta E(q) + \hbar \omega(Q))
                                                        \right] 
\left| <H_{pt}> \right|^2      \mbox{\hspace{3mm},}
\end{eqnarray}
where $ <H_{pt}>$ is the matrix element of the exciton recombination, 
which is proportional to $\delta_{n_h,n_e}$  and  
$\delta_{n_h,n_e-1} \sqrt{n_h}$ for the inter-band  and  intra-band
excitons, respectively, $n_e$ and $n_h$ denote the electron and the 
 hole Landau level numbers; $\Delta E(q)=E(q)-E(0)$,
 $\Delta \omega = \omega - \omega_0$, 
 $\omega _0$ is the photon frequency 
 of a direct optical emission; $N(q)$ and $f(q)$ are the phonon 
 and exciton distribution functions, respectively. 
 In expression (4) we took into account the conservation of
 momentum for the transitions into the intermediate state. 

 Below we consider two types of excitons:  the inter-band exciton 
with the electron and the hole being in the conduction and 
valence bands, respectively, and the intra-band intra-subband exciton 
with the electron and the hole being in the same size-quantization 
subband but in the different Landau levels. The hole in the later
case is the absence of the electron in the completely occupied 
Landau level.

For the inter-band exciton the optical transitions are allowed only if 
the electron and the hole have the same Landau level number. 
We study the case when they are both in the lowest Landau level. 
The exciton dispersion is given by the expression \cite{lerner,kallin}:
\begin{equation}
E(q) = \frac{e^2}{\kappa l} \int _{0}^{\infty }dk F(k) 
e^{-k^2l^2/2} J_{0}(kql^2)    \mbox{\hspace{3mm},}
\end{equation}
where $l$ is the magnetic length, $J_m$ is the Bessel 
function of the order $m$. The factor $F(k)$ is due to 
the finite width of the electron and hole wave functions 
in $z$-direction and has the form \cite{ando}:
\[
F(k) = \int\int dz_1 dz_2 e^{-k |z_1-z_2|} \left| \chi_e(z_1) \right|^2 
  \left| \chi_h(z_2) \right|^2    \mbox{\hspace{3mm}.}
\]
The matrix element of the electron (hole) density operator
 is given by the expression: $\rho (\vec{q}) = \exp(-q^2l^2/4)$.
We introduce the temperature $T$ of the system and assume 
that the exciton distribution function $f(q)$ is proportional to
 $ \exp (-E(q)/k_B T)$. 
Substituting equations (1)-(3),(5) into equation (4) we find the 
emission intensity, $I(\omega )$.  
The width of the emission line is given by the expression: 
$\delta \omega = \sqrt{ <\omega ^2> -<\omega >^2 }$,
where  $<\omega >$ and $<\omega ^2>$  are the first and
 the second normalized moments of the spectra, respectively.  
 In Fig.1(a) the typical emission line 
is shown for the width of the quantum well $a$ equal to 
the magnetic length $l$ and for 
 the temperature $T=2.5$K. The line has a single peak. 
In Fig.1(b) the width of the peak is shown as a 
function of temperature for different well widths. 
One can see that the width of the line is 
saturated with increasing temperature. This results from  the 
suppression of the phonon absorption (emission) for the values
of $q$ greater than the inverse magnetic length $1/l$. This 
 suppression is due to 
 the factor $\rho ^2(\vec{q}) = \exp(-q^2l^2/2)$ in the
 electron-phonon matrix elements. The magnetoexciton mass
 increases with increasing the well's width $a$. Therefore, 
 the width of the emission line decreases with increasing $a$.

For the intra-band exciton the shape of the emission line can be 
more complex. We consider the case when the electron filling 
factor is equal to three. 
The occupied levels are the first Landau level ($n=0$) 
with both spin directions, $S_z=+1/2$ and $S_z =-1/2$, and 
  the second Landau level ($n=1$) with spin $S_z = +1/2$. 
In this case the  optically active excitons are formed by 
the two types of excitations: 1) the hole is in the 
second Landau level ($n=1$, $S_z=+1/2$)  and the electron 
is in the third  Landau level ($n=2$, $S_z=+1/2$) and 
 2) the hole is in the first Landau level ($n=0$, $S_z=-1/2$)
 and the electron is in the second Landau level 
($n=1$, $S_z=-1/2$). These two types of  excitations 
 interact with each other. As a result of this interaction 
 (level repulsion) the exciton dispersion curve has a minimum 
 at non-zero momentum. The exciton dispersion for the lower branch
 is given by the expression \cite{kallin}:
\begin{equation}
E(q) = \frac{E_1(q)+E_2(q)}{2}-\frac{\sqrt{(E_1(q)-E_2(q))^2+4|V(q)|^2}}{2}
                     \mbox{\hspace{3mm},}
\end{equation}
where the energies $E_1(q)$, $E_2(q)$, and $V(q)$, written down
in units of $e^2/\kappa l$, are
\[
E_1(q) = \frac{q}{2}F(q)e^{-q^2/2}-
    \int _{0}^{\infty }dk F(k) \left(1-\frac{k^2}{2} \right) 
  e^{-k^2/2} J_{0}(kq)    \mbox{\hspace{3mm},}
\]
\[
E_2(q) = q\left(1-\frac{q^2}{4} \right)^2 F(q) e^{-q^2/2}-
    \int _{0}^{\infty }dk F(k) \left(1-\frac{k^2}{2} \right)
                               \left(1-k^2+\frac{k^4}{8} \right) 
    e^{-k^2/2} J_{0}(kq)   \mbox{\hspace{3mm},}
\]
and
\[
V(q) = \frac{q}{\sqrt{2}} \left(1-\frac{q^2}{4} \right)^2 F(q) e^{-q^2/2}
                           \mbox{\hspace{3mm},}  
\]
where the momentum is in units of $1/l$. 

The exciton dispersion is shown in Fig.2(a) for the two 
values of the well width. The dispersion has a minimum 
at a non-zero value of momentum. With increasing the well width 
the depth of the minimum decreases, which results from the 
decreasing of the effective electron-hole interaction. 
 The exciton emission line is shown in Fig.2(b). 
 At low temperature excitons occupy the non-zero 
 momentum minimum of the dispersion curve.
 Therefore there are only phonon-assisted optical 
transitions from this state. The optical line is red-shifted 
from the direct line, which is at $\Delta \omega =0$. 
With increasing temperature the population of the 
zero-momentum exciton state increases and an additional line, 
corresponding to the optical transition from the 
zero-momentum exciton, appears. 
The position of this line is at frequency $\omega _0$, which 
(according to the Kohn theorem \cite{kohn}) is equal to a
cyclotron frequency. At this intermediate temperature 
the exciton emission line has a double-peak structure. 
At higher temperature there is again a single line which now
 is at $\Delta \omega =0$. This line  corresponds to the 
transition from the zero-momentum exciton state without 
phonon emission or absorption. With increasing the well 
width the separation between the lines becomes smaller 
and the temperature, at which the lower energy peak can 
be observed, is decreasing.

 The additional peaks in the exciton absorption line were 
discussed in the literature \cite{kallin1,quinn} for the case 
of impurity scattering being the dominant mechanism of the 
momentum transfer. 

In conclusion, we have studied the phonon-assisted exciton 
emission. For the inter-band exciton the width of the 
emission line was calculated as a function of well width 
and temperature. The line-width is smaller for wider 
wells and is saturated with increasing temperature.
 For the intra-band exciton the double-peak structure of 
the emission line is predicted if the electron filling 
factor is equal to $2m+1$, where $m$ is an integer number. 
With increasing temperature the emission line acquires a 
blue shift.

This work was supported by the UK EPSRC.

\begin{figure}
\begin{center}
\begin{picture}(110,81)
\put(0,0){\includegraphics{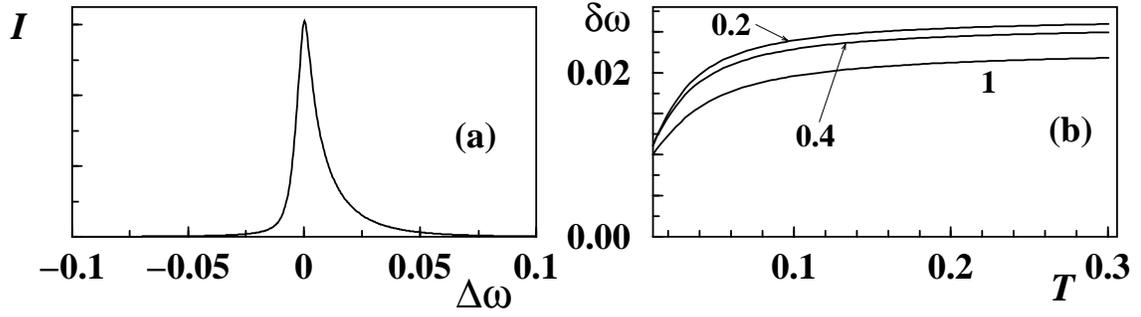}}
\end{picture}
\vspace*{2.0cm}
\caption{
(a) Inter-band exciton emission line for $a=l$ and $T=2.5$K. 
The intensity is in arbitrary units, $\Delta \omega $ is in 
units of $e^2/\kappa l$.
(b) The width of the exciton emission line as a function of temperature.
The numbers near the lines show the quantum well width in units of $l$. 
The temperature is in Kelvin. 
The line width is in units of $e^2/\kappa l$.
}
\end{center}
\end{figure}

\begin{figure}
\begin{center}
\begin{picture}(110,81)
\put(0,0){\includegraphics{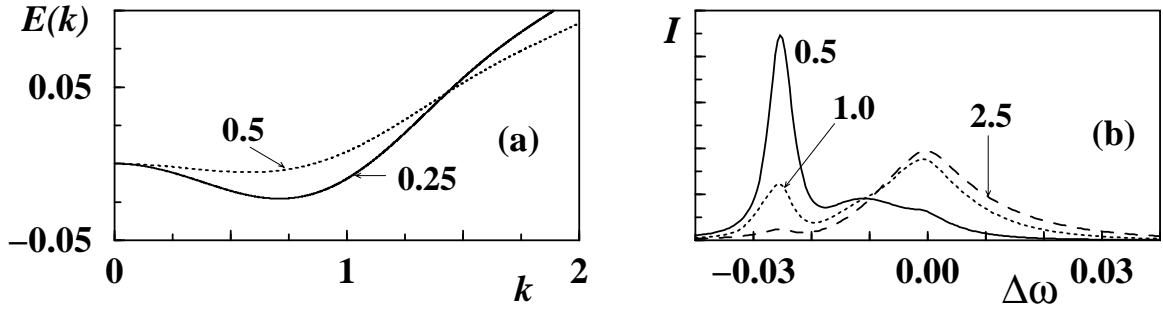}}
\end{picture}
\vspace*{2.0cm}
\caption{
(a) Intra-band exciton dispersion as a function of 
the momentum $k$. The energy $E(k)$ and the momentum $k$ are 
in units of $e^2/\kappa l$ and $1/l$, respectively.
 The numbers near the lines show the width of the quantum well in 
 units of magnetic length $l$. (b) Intra-band exciton 
 emission line at $a=0.25l$.  The numbers near the lines show
 the temperature in Kelvin. The intensity is in arbitrary 
 units and $\Delta \omega $ is in units of  $e^2/\kappa l$. 
}
\end{center}
\end{figure}

\end{document}